\documentclass{article}
\usepackage{spconf,amsmath,amssymb,graphicx}
\usepackage{physics}
\usepackage{nicefrac}
\usepackage{subcaption}
\usepackage{url}
\newcommand{\kronsum}{\raisebox{1pt}{\ensuremath{\:\oplus\:}}} % Kronceker sum

\usepackage{tikz,pgfplots}
\usetikzlibrary{plotmarks,shapes,arrows}
\pgfplotsset{compat=newest} 

\pgfplotsset{/pgf/number format/.cd, 1000 sep={}}

\pgfplotsset{every axis/.append style={
  grid style={line width=0.6pt,dotted,gray}}}

\pgfplotsset{every axis/.append style={
  legend style={inner xsep=1pt, inner ysep=0.5pt, nodes={inner sep=1pt, text depth=0.1em},draw=none,fill=none}
}}

\newlength\figureheight
\newlength\figurewidth

\title{Unifying Probabilistic Models for Time-Frequency Analysis}
\name{William J.\ Wilkinson$^1$\sthanks{Corresponding author: \texttt{w.j.wilkinson@qmul.ac.uk}.}, Michael Riis Andersen$^2$, Joshua D.\ Reiss$^1$, Dan Stowell$^1$, and Arno Solin$^2$\sthanks{AS acknowledges funding from the Academy of Finland (308640).}}

\address{%
  \hspace*{\fill}%
  \begin{minipage}{.4\textwidth}
  \centering
  $^1$Centre for Digital Music \\
  Queen Mary University of London \\
  United Kingdom
  \end{minipage}
  \hspace*{\fill}
  \begin{minipage}{.4\textwidth}
  \centering
  $^2$Department of Computer Science \\
  Aalto University \\
  Finland
  \end{minipage}%
  \hspace*{\fill}%
}

\begin{document}
%\ninept
%
\maketitle
\begin{abstract}
In audio signal processing, probabilistic time-frequency models have many benefits over their non-probabilistic counterparts. They adapt to the incoming signal, quantify uncertainty, and measure correlation between the signal's amplitude and phase information, making time domain resynthesis straightforward. However, these models are still not widely used since they come at a high computational cost, and because they are formulated in such a way that it can be difficult to interpret all the modelling assumptions. By showing their equivalence to Spectral Mixture Gaussian processes, we illuminate the underlying model assumptions and provide a general framework for constructing more complex models that better approximate real-world signals. Our interpretation makes it intuitive to inspect, compare, and alter the models since all prior knowledge is encoded in the Gaussian process kernel functions. We utilise a state space representation to perform efficient inference via Kalman smoothing, and we demonstrate how our interpretation allows for efficient parameter learning in the frequency domain.

\end{abstract}

\section{Introduction}
\label{sec:intro}
\begin{figure}[t]
	%\centering
	%\tikz\node[shape=rectangle,draw=gray,fill=gray!10,minimum width=.48\textwidth, minimum height=.48\textwidth] {(figure goes here)};
	\centering\scriptsize
	\pgfplotsset{yticklabel style={rotate=90}, ylabel style={yshift=0pt},clip=true,scale only axis,axis on top,clip marker paths, title style={yshift=-4pt, font=\footnotesize}, xlabel style={font=\footnotesize}}
	\setlength{\figurewidth}{.44\columnwidth}
	\setlength{\figureheight}{.7\figurewidth}
	\hspace*{-3em}
	\begin{subfigure}[b]{.48\columnwidth}
	    \raggedleft
		\input{./fig/spec_density.tex}
		%\caption{...}
	\end{subfigure}
	%
	%\hspace*{\fill}
	\hspace*{2em}
	\begin{subfigure}[b]{.48\columnwidth}
	    \raggedleft	
		\input{./fig/kernels.tex}
		%\caption{...}
	\end{subfigure}  
	\\
	\hspace*{-3em}
	\begin{subfigure}[b]{.48\columnwidth}
	    \raggedleft	
\begin{tikzpicture}

\begin{axis}[%
point meta min=-0.25,
point meta max=0.5,
axis on top,
xmin=0.5,
xmax=3000.5,
xtick={500,1500,2500},
xticklabels={{$\nu=\nicefrac{1}{2}$},{$\nu=\nicefrac{3}{2}$},{$\nu=\nicefrac{5}{2}$}},
xlabel={\phantom{FOO}},
ymin=0.5,
ymax=3000.5,
ytick={300,1500,2700},
yticklabels={{$\nu=\nicefrac{5}{2}$},{$\nu=\nicefrac{3}{2}$},{$\nu=\nicefrac{1}{2}$}},
ylabel={\phantom{FOO}},
axis background/.style={fill=white},
title={Covariance matrices},
legend style={legend cell align=left,align=left,draw=white!15!black},
width=\figurewidth,
height=\figureheight
]
\addplot [forget plot] graphics [xmin=0.5,xmax=3000.5,ymin=0.5,ymax=3000.5] {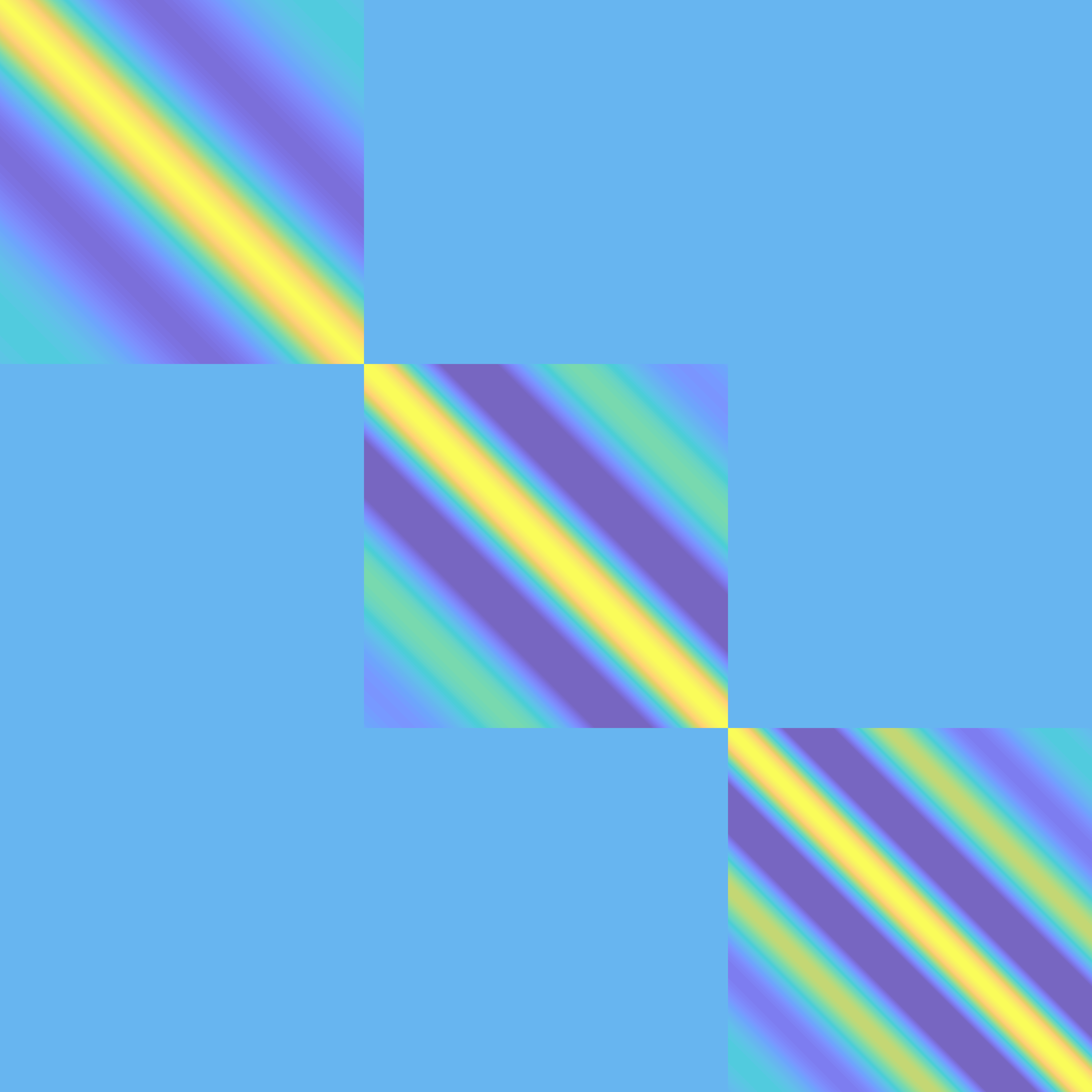};
\end{axis}
\end{tikzpicture}%
		%\caption{...}
	\end{subfigure}
	%
	%\hspace*{\fill}
	\hspace*{2em}
	\begin{subfigure}[b]{.48\columnwidth}
	    \raggedleft	
		\input{./fig/samples.tex}
		%\caption{...}
	\end{subfigure} 
	\\[-1em]
	\caption{Four representations of the same Gaussian process-based probabilistic filter bank (with three filters). Each filter / process is a frequency shifted Mat\'ern-$\nu$ GP. All three filters have the same lengthscale (bandwidth) parameter, but they exhibit quite different spectral densities ({\bf top-left}). See main text for demonstration of how filter banks can be represented in canonical GP form, such as with kernel functions ({\bf top-right}) and covariance matrices ({\bf bottom-left}). Samples from the prior vary in smoothness ({\bf bottom-right}), suggesting that the choice of $\nu$ will affect how the model fits the signal.}
	\label{fig:audio}
	\vspace{-1em}
\end{figure}
Time-frequency (TF) analysis is a ubiquitous technique for uncovering the time-varying spectral properties of signals, and it commonly plays the part of a pre-processing module for machine learning and signal processing tasks. However, traditional TF analysis requires various choices to be made regarding windowing functions, transfer functions or wavelets, depending on the representation being used \cite{leon1995time}. It is not clear how best to make these choices or what their implications are on tasks such as classification or source separation.

Probabilistic TF analysis \cite{turner-time14} promises to remove the need for these difficult decisions by adapting to the incoming signal \cite{sejdic2009time,qi2002bayesian,cemgil2005probabilistic,zhong2010time} and by propagating uncertainty information to downstream applications \cite{turner-time14,gillespie2001optimizing}. By specifying a probabilistic model characterised by parameters corresponding to traditional model features, such as centre frequencies and bandwidths of a filter bank, a posterior distribution over the frequency components given the data can be found. Different modelling choices can be compared in a principled manner by evaluating the model likelihood given the parameters, which allows for parameter tuning in order to find the statistically optimal TF representation for a given signal.

Probabilistic models that act directly on the signal waveform implicitly measure correlation between a signal's amplitude and phase information \cite{turner2010statistical}, which has the major implication that time-domain synthesis does not require a phase-reconstruction stage. This ability to sample new data from the generative model makes missing data imputation and noise reduction tasks intuitive.

Despite these benefits, existing probabilistic TF models are still not widely used, perhaps due to their higher computational complexity and because they are formulated in such a way that they can be difficult to interpret and understand.

In the field of machine learning, Gaussian processes (GPs) \cite{rasmussen2006gaussian} are an increasingly popular non-parametric approach for learning and decision making. In their standard formulation, they are characterised by covariance matrices that capture the hidden structure of data. Covariance matrices are constructed by evaluating \emph{kernel functions} that encode our prior knowledge about the system we are modelling. A major issue with this approach for time-series data is that evaluation of the covariance matrix is impractical for all but the shortest of real-world signals. It is well known that many probabilistic TF methods can be posed as GPs \cite{turner-time14}, but it is generally unclear how all the modelling assumptions relate to the standard set of GP techniques.

In \cite{wilson2013gaussian}, GPs, along with their neural network counterparts, are presented as \emph{``intelligent agents''} capable of automating the learning and decision making process. It is shown how complex prior knowledge can be encoded in the system by constructing new kernel functions composed of the sum and product of simpler ones. One such class of functions presented in \cite{wilson2013gaussian} are Spectral Mixture kernels, defined for one-dimensional inputs as
\vspace{-0.2cm}
$$\kappa_{\mathrm{sm}}(t,t') = \sum_{d=1}^D\sigma^2_d\cos(\omega_d(t-t'))\exp(-(t-t')^2/\ell_d^2),\vspace{-0.2cm}$$
which comprises a sum of frequency-shifted \emph{radial basis function} kernels \cite{rasmussen2006gaussian} and whose spectral density is a mixture of Gaussians. This idea is extended to the entire Mat\'ern kernel class in \cite{alvarado2017efficient, alvarado2018sparse}, producing Cauchy-Lorentz densities.

In this work, we show that probabilistic TF analysis and Mat\'ern Spectral Mixture GPs are in fact equivalent. In other words, Spectral Mixture kernels are probabilistic filter banks. By doing so we reinterpret TF modelling assumptions under the GP paradigm. We provide a general procedure for rewriting Spectral Mixture GPs in discrete state space form, such that more complex TF models can be easily constructed, and inference can be performed efficiently via Kalman smoothing, whose computational complexity scales linearly in the number of time steps $T$ and cubicly in state dimensionality $M$, $\mathcal{O}(M^3T)$. We then show how to utilise the continuous spectral density of the kernel functions to optimise the model parameters in the frequency domain.\footnote{Matlab code for all methods and experiments is available at:\\ \url{https://github.com/wil-j-wil/unifying-prob-time-freq}}

After outlining our framework and formalising the equivalence between these two modelling paradigms in Section~\ref{sec:model}, we go on to illustrate some potential modifications to the standard probabilistic TF approach in Section~\ref{sec:experiments}, evaluating the impact of these updates on a missing data synthesis task.

\section{State space Gaussian process models for time-frequency analysis}
\label{sec:model}
Various models for Bayesian treatment of time-frequency analysis have been proposed, most notably Bayesian spectrum estimation (BSE) \cite{qi2002bayesian}, and the probabilistic phase vocoder (PPV) \cite{cemgil2005probabilistic}. For an overview see \cite{turner-time14}, where it is also shown that the BSE and PPV models are equivalent up to a shift in frequency. We proceed by considering the PPV model, reformulating it with the aim of unifying these models in a common Gaussian process framework to illuminate the underlying modelling assumptions.

\emph{Readers should keep in mind that all the models \eqref{eq:PPV}--\eqref{eq:SS_disc_2} written in this section are exactly equivalent to one another. By presenting them this way, we show multiple perspectives on spectral data analysis.}

\subsection{Probabilistic phase vocoder}
The standard discrete-time PPV can be written as follows,
\begin{gather}
	\begin{aligned}
		&\text{[Prior]}  & \mathrm{z}_{d,k} &= \psi_d \mathrm{e}^{\mathrm{i}\,\omega_d} \mathrm{z}_{d, k-1} + \rho_d \, \zeta_{d,k}, \\
		&\text{[Likelihood]}  & {y}_k &= \sum_{d=1}^D \Re[\mathrm{z}_{d,k}] + \sigma_{\mathrm{y}_k}\varepsilon_k, 
	\end{aligned} \label{eq:PPV}
\end{gather}
where $k$ indexes the time step, with complex phasor $z_{d,k} \in \mathbb{C}$ being the (latent) subband signal in frequency channel $d = 1,\ldots,D$. $y_k$ denotes the observed audio signal at $t_k$ and both $\varepsilon_k$ and $\zeta_{d,k}$ are i.i.d.\ Gaussian noise $\mathrm{N}(0,1)$, real-valued and complex-valued, respectively. Note that the noise scale $\sigma_{\mathrm{y}_k}$ can be non-stationary. Parameters $\psi_d$ and $\rho_d$ represent the process and noise variances respectively, whilst $\omega_d$ is the instantaneous angular frequency.

Recognising that Eq.~\eqref{eq:PPV} is a complex first-order autoregressive process makes it straightforward to write down the model's state space form,
\begin{gather}
	\begin{aligned}
		&\text{[Prior]}  & \mathbf{z}_{k+1} &= \mathbf{A}\mathbf{z}_k + \mathbf{q}_k, & \hspace{-0.1cm} \mathbf{q}_k\sim \mathrm{N}(\mathbf{0},\mathbf{Q}),\\
		&\text{[Likelihood]}  & {y}_k &= \mathbf{H}\mathbf{z}_k + \sigma_{\mathrm{y}_k}\varepsilon_k,&
	\end{aligned} \label{eq:SS_disc}
\end{gather}
for $\mathbf{z}_k=\left(\Re[\mathrm{z}_{1,k}]~\Im[\mathrm{z}_{1,k}]~\hdots~\Re[\mathrm{z}_{D,k}]~\Im[\mathrm{z}_{D,k}]\right)^\mathsf{T}$ and measurement matrix $\mathbf{H}=(1~0~\hdots~1~0)$, with transition matrix $\mathbf{A}$ and process noise covariance matrix	 $\mathbf{Q}$ defined by
\begin{align*}
&& \mathbf{A} = \left(\begin{smallmatrix} \psi_1\mathbf{R}(\omega_1) &  & \makebox(0,0){\text{\Large0}} \\  & \ddots & \\ \makebox(10,10){\text{\Large0}} & & \psi_D\mathbf{R}(\omega_D) \end{smallmatrix}\right), && \mathbf{Q} = \left(\begin{smallmatrix} \rho_1^2\mathbf{I} &  & \makebox(0,0){\text{\Large0}} \\  & \ddots & \\ \makebox(10,10){\text{\Large0}} & & \rho_D^2\mathbf{I} \end{smallmatrix}\right), &&
\end{align*}
for rotation matrix $\mathbf{R}(\omega_d)=\left(\begin{smallmatrix}\cos\omega_d&-\sin\omega_d\\
\sin\omega_d & \cos\omega_d\end{smallmatrix}\right)$.

This linear Gaussian dynamical system is in the precise form required for inference via Kalman filtering and smoothing \cite{sarkka2013bayesian}. The filtering equations provide us with the necessary information required to evaluate the marginal likelihood $p(\mathbf{y}|\theta)$ and hence perform hyperparameter tuning. However, in practice it is much more efficient to tune the hyperparameters in the frequency domain, as discussed in the Section \ref{sub:opt}.

\subsection{PPV model in canonical GP form}
Gaussian processes are commonly used in Bayesian inference as non-parametric prior distributions on functions \cite{rasmussen2006gaussian}. A GP prior, $f \sim \mathrm{GP}\left(\mu(\cdot), \kappa(\cdot, \cdot)\right)$, is completely characterized by a mean function, $\mu(t)$, and a kernel function, $\kappa(t,t')$. 

We first write down the PPV's kernel-based GP representation, before going on to show equivalence to Eq.~\eqref{eq:PPV},
\begin{gather}
	\begin{aligned}
		&\text{[Prior]}   &\hspace{-0.1cm} f(t) &\sim \mathrm{GP}(0, \sum_{d=1}^D \kappa^{(d)}_{\mathrm{cos}}(t,t')\,\kappa^{(d)}_{\mathrm{exp}}(t,t')), \\
		&\text{[Likelihood]} \!& {y}_k &= f(t_k) + \sigma_{\mathrm{y}_k}\,\varepsilon_k, 
	\end{aligned} \label{eq:GP_model}
\end{gather}
where $\kappa^{(d)}_{\mathrm{cos}}(t,t') = \cos(\omega_d\,(t-t'))$ is a deterministic kernel whose function realisations are pure sinusoids, and $\kappa^{(d)}_{\mathrm{exp}}(t,t') = \sigma^2_d\exp(-|t-t'|/\ell_d)$ is the exponential kernel, otherwise known as the Mat\'ern-$\nicefrac{1}{2}$. The cosine kernel acts as a frequency shift operator, centring the spectral density of the exponential kernel around $\omega_d$.

When written in this form, it becomes apparent that the kernel in Eq.~\eqref{eq:GP_model} has a close connection with Spectral Mixture kernels. Specifically it belongs to the Mat\'ern Spectral Mixture family used to model harmonic priors over musical audio signals \cite{alvarado2017efficient,alvarado2018sparse}. This explicit link with Spectral Mixture models has not been previously explored, however filter banks composed in this way are reminiscent of the exponential kernels described in \cite{choi1989improved}.%, so in some sense these GP-based models are a return to classical ideas, updated to exploit modern inference methods.

We will now demonstrate the equivalence between the model in Eq.~\eqref{eq:GP_model} and the PPV model in Eq.~\eqref{eq:SS_disc} by converting it back to discrete state space form. In doing so, we will outline a general procedure for reformulating models in the form of Eq.~\eqref{eq:GP_model} in this way, such that efficient inference methods can be applied even after the model has been altered.

\subsection{The corresponding continuous state space model}
The model in Eq.~\eqref{eq:GP_model} has an equivalent representation in terms of a linear time-invariant (LTI) stochastic differential equation (SDE, see, e.g., \cite{Sarkka+Solin:inpress}). The prior (or dynamics) of the system can be written in terms of a driving Brownian motion (with spectral density $\mathbf{Q}_\mathrm{c}$) in It\^o form:
\begin{gather}
	\begin{aligned}
		&\text{[Prior]}   & \mathrm{d}\mathbf{f}(t) &= \mathbf{F}\mathbf{f}(t)\,\mathrm{d}t + \mathbf{L}\,\mathrm{d}\boldsymbol{\beta}(t),&  \\
		&\text{[Likelihood]} \!&  y_k &= \mathbf{\tilde{H}}\mathbf{f}(t_k) + \sigma_{\mathrm{y}_k}\varepsilon_k,&
	\end{aligned}\label{eq:SS_cts}
\end{gather}
where $\mathbf{f}(t) : \mathbb{R} \to \mathbb{R}^{M}$ for state space order $M$, and $\mathbf{F}$, $\mathbf{L}$, and $\mathbf{\tilde{H}}$ are model matrices. Building on previous work \cite{pmlr-v33-solin14}, we may write the GP prior in Eq.~\eqref{eq:GP_model} as an LTI SDE of block-Kronecker structure:
\begin{align*}
 \mathbf{F} = \mathrm{blkdiag}(\mathbf{F}_\mathrm{cos}^{(1)} \kronsum \mathbf{F}_\mathrm{exp}^{(1)}, \ldots, \mathbf{F}_\mathrm{cos}^{(D)} \kronsum \mathbf{F}_\mathrm{exp}^{(D)}),
\end{align*}
where `$\!\kronsum\!$' denotes the Kronecker sum, $\mathbf{F}_\mathrm{cos}^{(d)} = \left(\begin{smallmatrix}0 & -\omega_d \\ \omega_d & 0 \end{smallmatrix}\right)$, and $\mathbf{F}_\mathrm{exp}^{(d)} = -1/\ell_d$ (because the exponential covariance function has an exact LTI SDE representation \cite{Sarkka+Solin:inpress}). Similarly, the rest of the model matrices are given in terms of the Kronecker products of the submodel matrices.

\subsection{Returning to discrete state space form} \label{sub:disc_model}
LTI SDE models such as Eq.~\eqref{eq:SS_cts} have an exact discrete-time solution, and the corresponding state space model is given by \cite{Sarkka+Solin:inpress}:
\begin{gather}
	\begin{aligned}
		&\text{[Prior]}   & \mathbf{f}_{k+1} &= \mathbf{\tilde{A}}\mathbf{f}_k + \mathbf{\tilde{q}}_k,& \mathbf{\tilde{q}}_k &\sim \mathrm{N}(\mathbf{0},\mathbf{\tilde{Q}}),  \\
		&\text{[Likelihood]} \!&  y_k &= \mathbf{\tilde{H}}\mathbf{f}_k +  \sigma_{\mathrm{y}_k}\varepsilon_k,&  &
	\end{aligned} \label{eq:SS_disc_2}
\end{gather}
where $\mathbf{f}_k$ is the $M$ dimensional state, $\mathbf{\tilde{A}} = \exp(\mathbf{F}\,\Delta t)$ and $\mathbf{\tilde{Q}} = \mathbf{P_{\infty}}-\mathbf{\tilde{A}}\mathbf{P_{\infty}}\mathbf{\tilde{A}}^\mathsf{T}$. The stationary state covariance $\mathbf{P_{\infty}}$ is straightforward to calculate for most common kernel functions \cite{Sarkka+Solin:inpress}, and in the exponential kernel case is $\mathbf{P_{\infty}}=\mathrm{diag}(\sigma^2_1,\hdots,\sigma^2_D)$.

Given that ${\exp}(\mathbf{F}_\mathrm{cos}^{(d)}\Delta t)=\mathbf{R}(\omega_d)$, performing these calculations for our PPV model ($M=2D$) results in the following parametrisation:
\begin{align*}
&& \mathbf{\tilde{A}}=\left(\begin{smallmatrix} \eta_1\mathbf{R}(\omega_1) &  & \makebox(0,0){\text{\Large0}} \\  & \ddots & \\ \makebox(10,10){\text{\Large0}} & & \eta_D\mathbf{R}(\omega_D) \end{smallmatrix}\right), && \mathbf{\tilde{Q}} = \left(\begin{smallmatrix} \alpha_1\mathbf{I} &  & \makebox(0,0){\text{\Large0}} \\  & \ddots & \\ \makebox(10,10){\text{\Large0}} & & \alpha_D\mathbf{I} \end{smallmatrix}\right), &&
\end{align*}
where $\eta_d=\exp(-\Delta t/\ell_d)$ and $ \alpha_d=\sigma^2_d(1-\exp(-2\Delta t/\ell_d))$. Finally it is clear that the models in Eq.~\eqref{eq:SS_disc_2} and Eq.~\eqref{eq:SS_disc} are of identical form, and hence the probabilistic phase vocoder in Eq.~\eqref{eq:PPV} is equivalent to the GP model in Eq.~\eqref{eq:GP_model} if we select parameters $\psi_d=\exp(-\Delta t/\ell_d)$ and $\rho^2_d=\sigma^2_d(1-\exp(-2\Delta t/\ell_d))$.

Whilst we have derived this model for the exponential kernel, the framework outlined above can be followed regardless of the kernel choice, as long as it can be written in state space form. This is possible for most commonly used kernel functions \cite{Sarkka+Solin:inpress, hartikainen2010kalman}. An intuitive way to proceed now is to alter the kernel in Eq.~\eqref{eq:GP_model} to fit our requirements for the form of the corresponding filter bank. We explore this idea in Section \ref{sec:experiments}.

\subsection{Frequency domain optimisation}\label{sub:opt}

The formal connection to probabilistic TF models allows us to utilise Bayesian spectrum analysis \cite{bretthorst2013bayesian,turner-time14} for frequency domain hyperparameter tuning in Spectral Mixture GPs. This is significantly faster than time domain optimisation, and we avoid getting stuck in local optima by fitting to a smoothed version of the signal spectrum. By optimising parameters of a stationary GP kernel, rather than coefficients of an auto-regressive process, we guarantee stationarity of the filter bank. The Spectral Mixture GP perspective gives us direct access to the model spectrum $\gamma_{y,i}(\theta)$ via the sum of the kernel's spectral density functions, $\gamma_{y,i}(\theta)=\sum_{d=1}^D  S_{d,i}(\theta)+T\sigma^2_y$, where $S_{d,i}(\theta)$ is the spectral density of the kernel $\kappa^{(d)}$ in Eq.~\eqref{eq:GP_model} evaluated at frequency bin $i$.

We fit the parameters via optimisation of the log-likelihood,
\begin{align*}
\log p(y|\theta)=c-\frac{1}{2}\sum_{i=1}^T\left(  \log(\gamma_{y,i}(\theta))+\frac{|\tilde{y}_i|^2}{\gamma_{y,i}(\theta)} \right),
\end{align*}
where $|\tilde{y}_i|^2=|\sum_{k=1}^T \mathrm{FT}_{i,k}y_k|^2$ is the signal spectrum. Note that the cosine kernel in Eq.~\eqref{eq:GP_model} shifts the spectral density of the exponential kernel such that $S_{d,i}=\frac{1}{2}\left(S^{\mathrm{exp}}_{d,i-\omega_d}+S^{\mathrm{exp}}_{d,i+\omega_d}\right)$.

\section{Missing Data Experiment}
\label{sec:experiments}

The methodology outlined in Section \ref{sec:model} allows new TF models to be constructed, with increased freedom over the choice of covariance structure. Here we demonstrate the potential benefits of altering the modelling assumptions via a missing data synthesis task, similar to the one carried out in \cite{turner-time14}.

The first-order state space form of standard TF models (Eq.~\eqref{eq:PPV}) implies that instantaneous frequencies are not correlated through time \cite{turner2010statistical}. Higher-order models encourage slowly-varying instantaneous frequencies, a feature of real-world signals that should be leveraged to aid the highly ill-posed task of inferring a TF representation from data.

Therefore one intuitive example of a way to update the model is to swap the exponential (Mat\'ern-$\nicefrac{1}{2}$, state dimensionality $M=2D$) kernel with a similar function that admits a higher-order state space representation. This corresponds to a filter bank whose filter transfer functions are no longer first-order autoregressive processes, but take a more complex form. We use the Mat\'ern-$\nicefrac{3}{2}$ ($M=4D$) and Mat\'ern-$\nicefrac{5}{2}$ ($M=6D$) kernels, which correspond to second- and third-order filter banks respectively and whose spectral densities have flatter tails and taller peaks (see Fig.~\ref{fig:audio}). Note that the Mat\'ern-$\nicefrac{1}{2}$ model corresponds to the standard PPV.

\begin{figure}[t]
  %\centering
  %\tikz\node[shape=rectangle,draw=gray,fill=gray!10,minimum width=.48\textwidth, minimum height=.48\textwidth]
  \centering\scriptsize
  \pgfplotsset{yticklabel style={rotate=90}, ylabel style={yshift=0pt},clip=true,scale only axis,axis on top,clip marker paths, title style={yshift=-6pt, font=\small}, xlabel style={font=\small}}
  \setlength{\figurewidth}{.43\columnwidth}
  \setlength{\figureheight}{1.1\figurewidth}
  \begin{subfigure}[b]{.48\columnwidth}
\definecolor{mycolor1}{rgb}{0.90000,0.40000,0.20000}%
\definecolor{mycolor2}{rgb}{0.30000,0.50000,0.80000}%
\definecolor{mycolor3}{rgb}{0.40000,0.80000,0.20000}%
\begin{tikzpicture}

\begin{axis}[%
xmin=0,
xmax=20,
xlabel style={font=\color{white!15!black}},
xlabel={missing data duration (ms)},
axis background/.style={fill=white},
title={PESQ},
legend style={legend cell align=left, align=left, draw=white!15!black},
width=\figurewidth,
height=\figureheight
]

\addplot[area legend, draw=black, fill=mycolor1, draw opacity=0, fill opacity=0.15]
table[row sep=crcr] {%
x	y\\
0.625	4.00429839705428\\
3.875	3.47747071334639\\
7.125	3.31868034924753\\
10.3125	2.97127679475023\\
13.5625	2.61842836374345\\
16.8125	2.33424993819639\\
20	2.04333934410384\\
20	2.36665456812717\\
16.8125	2.62505163016455\\
13.5625	2.83640656238792\\
10.3125	3.14503350227754\\
7.125	3.46577842341083\\
3.875	3.60571552360066\\
0.625	4.16050030632437\\
}--cycle;

\addplot[area legend, draw=black, fill=mycolor2, draw opacity=0, fill opacity=0.15]
table[row sep=crcr] {%
x	y\\
0.625	4.08052523197377\\
3.875	3.53488778662807\\
7.125	3.38096681516509\\
10.3125	3.02160066343772\\
13.5625	2.66640166442658\\
16.8125	2.32517405048441\\
20	1.98653980420256\\
20	2.3054778404993\\
16.8125	2.60578584347943\\
13.5625	2.88431200161642\\
10.3125	3.18364463995376\\
7.125	3.55664268557344\\
3.875	3.68354086482907\\
0.625	4.23230768449141\\
}--cycle;

\addplot[area legend, draw=black, fill=mycolor3, draw opacity=0, fill opacity=0.15]
table[row sep=crcr] {%
x	y\\
0.625	4.05124682775178\\
3.875	3.61071188900614\\
7.125	3.28924240596559\\
10.3125	3.02919020714269\\
13.5625	2.61074468058386\\
16.8125	2.30567559200086\\
20	1.93940472808622\\
20	2.28157746967297\\
16.8125	2.58200757164142\\
13.5625	2.84836795283376\\
10.3125	3.20276942742106\\
7.125	3.43449660219666\\
3.875	3.73046078439424\\
0.625	4.1752497186699\\
}--cycle;

\addplot [color=mycolor1, line width=0.7pt]
  table[row sep=crcr]{%
0.625	4.08239935168932\\
3.875	3.54159311847352\\
7.125	3.39222938632918\\
10.3125	3.05815514851388\\
13.5625	2.72741746306568\\
16.8125	2.47965078418047\\
20	2.2049969561155\\
};

\addplot [color=mycolor2, line width=0.7pt]
  table[row sep=crcr]{%
0.625	4.15641645823259\\
3.875	3.60921432572857\\
7.125	3.46880475036926\\
10.3125	3.10262265169574\\
13.5625	2.7753568330215\\
16.8125	2.46547994698192\\
20	2.14600882235093\\
};

\addplot [color=mycolor3, line width=0.7pt]
  table[row sep=crcr]{%
0.625	4.11324827321084\\
3.875	3.67058633670019\\
7.125	3.36186950408112\\
10.3125	3.11597981728188\\
13.5625	2.72955631670881\\
16.8125	2.44384158182114\\
20	2.1104910988796\\
};

\end{axis}
\end{tikzpicture}%
    %\caption{...}
  \end{subfigure}
  \hspace*{\fill}
  \begin{subfigure}[b]{.48\columnwidth}
\definecolor{mycolor1}{rgb}{0.90000,0.40000,0.20000}%
\definecolor{mycolor2}{rgb}{0.30000,0.50000,0.80000}%
\definecolor{mycolor3}{rgb}{0.40000,0.80000,0.20000}%
\begin{tikzpicture}

\begin{axis}[%
xmin=0,
xmax=20,
xlabel style={font=\color{white!15!black}},
xlabel={missing data duration (ms)},
axis background/.style={fill=white},
title={SNR (dB)},
legend style={legend cell align=left, align=left, draw=white!15!black},
width=\figurewidth,
height=\figureheight
]

\addplot[area legend, draw=black, fill=mycolor1, draw opacity=0, fill opacity=0.15, forget plot]
table[row sep=crcr] {%
x	y\\
0.625	7.32868091230665\\
3.875	4.55050527143606\\
7.125	3.19098710923508\\
10.3125	2.62876470049576\\
13.5625	1.80378357299206\\
16.8125	1.102361337004\\
20	1.02507347116123\\
20	2.03851361319748\\
16.8125	2.16800710175545\\
13.5625	2.84284671018187\\
10.3125	3.72004917452155\\
7.125	4.60282890811198\\
3.875	5.93647950854414\\
0.625	9.94892918842559\\
}--cycle;

\addplot[area legend, draw=black, fill=mycolor2, draw opacity=0, fill opacity=0.15, forget plot]
table[row sep=crcr] {%
x	y\\
0.625	7.83877829878361\\
3.875	5.12065277271947\\
7.125	3.8535220893063\\
10.3125	3.27739992927271\\
13.5625	2.0535638510769\\
16.8125	1.73568947614987\\
20	1.5579107085697\\
20	2.5269973922696\\
16.8125	2.69133009480306\\
13.5625	3.00114867523941\\
10.3125	4.24154511634815\\
7.125	5.23043947685728\\
3.875	6.52470058333841\\
0.625	11.0188984022382\\
}--cycle;

\addplot[area legend, draw=black, fill=mycolor3, draw opacity=0, fill opacity=0.15, forget plot]
table[row sep=crcr] {%
x	y\\
0.625	9.38969045745871\\
3.875	5.14443902861141\\
7.125	3.66124854686384\\
10.3125	3.68844872955864\\
13.5625	2.87054652338456\\
16.8125	1.99951700651334\\
20	1.42963209844509\\
20	2.62262558122171\\
16.8125	3.35291509128876\\
13.5625	4.18794115155408\\
10.3125	5.08180619977239\\
7.125	5.44315220403948\\
3.875	6.75877900567393\\
0.625	12.2755505524926\\
}--cycle;
\addplot [color=mycolor1, line width=0.7pt]
  table[row sep=crcr]{%
0.625	8.63880505036612\\
3.875	5.2434923899901\\
7.125	3.89690800867353\\
10.3125	3.17440693750866\\
13.5625	2.32331514158697\\
16.8125	1.63518421937973\\
20	1.53179354217935\\
};
\addlegendentry{$\nu=\nicefrac{1}{2}$}

\addplot [color=mycolor2, line width=0.7pt]
  table[row sep=crcr]{%
0.625	9.42883835051092\\
3.875	5.82267667802894\\
7.125	4.54198078308179\\
10.3125	3.75947252281043\\
13.5625	2.52735626315815\\
16.8125	2.21350978547646\\
20	2.04245405041965\\
};
\addlegendentry{$\nu=\nicefrac{3}{2}$}

\addplot [color=mycolor3, line width=0.7pt]
  table[row sep=crcr]{%
0.625	10.8326205049757\\
3.875	5.95160901714267\\
7.125	4.55220037545166\\
10.3125	4.38512746466552\\
13.5625	3.52924383746932\\
16.8125	2.67621604890105\\
20	2.0261288398334\\
};
\addlegendentry{$\nu=\nicefrac{5}{2}$}

\end{axis}
\end{tikzpicture}%
    %\caption{...}
  \end{subfigure}  
  \\[4pt]
  \setlength{\figureheight}{.8\figurewidth}    
  \setlength{\figurewidth}{.95\columnwidth}
  \hspace*{\fill}%
  \begin{subfigure}[b]{\columnwidth}
    \input{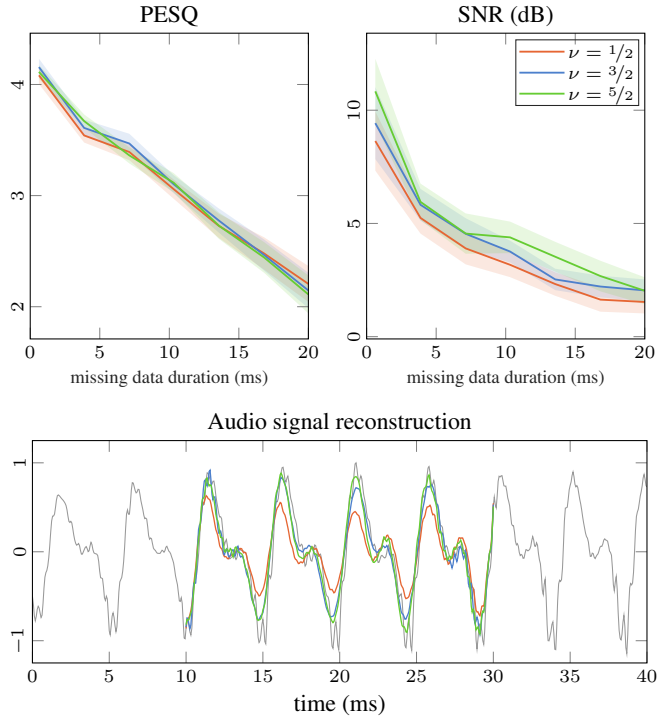}
    %\caption{...}
  \end{subfigure}    
  \hspace*{\fill}
  \vspace*{-1.5em}%
  \caption{Missing data synthesis results for three Mat\'ern-$\nu$ probabilistic time-frequency models. Segments of data were removed from 10 speech recordings. Performance measured via perceptual quality metric ({\bf top-left}) and signal-to-noise ratio ({\bf top-right}) as a function of gap duration. Median value across speakers shown (shaded area is standard error). A reconstruction example ({\bf bottom}) shows how the higher-order models ($\nu=\nicefrac{3}{2},\nicefrac{5}{2}$) recover the overall shape in clearer detail (ground truth in grey). Mat\'ern-$\nicefrac{1}{2}$ is the standard probabilistic phase vocoder.}
  \label{fig:results}
\end{figure}

Each model, with $D=40$ filters, was trained on 10 short speech excerpts (between 1 and 2~seconds in duration) and then used to filter versions of the recordings in which some data had been removed. Missing data gaps of between 1~ms and 20~ms were studied, with the results shown in Fig.~\ref{fig:results}. Whilst the differences are subtle (the overall models are similar), the higher-order models' reconstruction achieved an improved signal to noise ratio for all missing data durations averaged across the 10 speakers. We also calculated the PESQ score \cite{rix2001perceptual} (a standardised perceptual speech quality metric), which demonstrated some signs of improvement, however all models performed similarly for large gap durations.

\section{Discussion}
\label{sec:summary}
This paper serves to unify the theory surrounding probabilistic time-frequency analysis and explain clearly how it relates to Gaussian process modelling, with the hope of motivating further research at the intersection of these fields. We provide a general framework for converting spectral mixture GP models to a state space form that enables efficient frequency domain optimisation and efficient time domain filtering and prediction. We applied the framework to Mat\'ern Spectral Mixture GPs and demonstrated improved performance over the standard probabilistic phase vocoder on a generative task.

Practical limitations of probabilistic time-frequency models still remain due to the Kalman smoother's cubic computational scaling in the state dimensionality and from the significant memory requirements involved in storing the entire covariance structure for every time step. Future work must focus on these practical issues.

Importantly, the methods presented here assume independence across frequency channels and don't explicitly model time-varying amplitude behaviour. It has been shown previously that a joint model over the TF representation and the amplitudes can result in significant improvement on tasks such as synthesis and noise reduction. Our state space framework provides a foundation on which to construct these more complex models.

% -------------------------------------------------------------------------
\begingroup
\bibliographystyle{IEEEbib}
\bibliography{refs}
\endgroup
\end{document}